\documentclass{aa}
\usepackage{graphicx}
\usepackage{natbib}

\begin{document}
   \title{A close look into the carbon disk at the core of the planetary nebula CPD-56$^\circ$8032\thanks{Based on observations made with the Very Large Telescope Interferometer at
       Paranal Observatory under programs 073.D-0130}}

\titlerunning{The disk of CPD-568032}

\authorrunning{Chesneau, O. et al.} 

   \author{O.~Chesneau\inst{1}, A.~Collioud\inst{1}, O.~De~Marco\inst{2}, S.~Wolf\inst{3}, E.~Lagadec\inst{4}, A.A.~Zijlstra\inst{5}, 
   A.~Rothkopf\inst{3}, A.~Acker\inst{6}, G.C.~Clayton\inst{7}, B.~Lopez\inst{1}}

   \offprints{O. Chesneau}

\institute{Observatoire de la C\^{o}te d'Azur, Dpt. Gemini-CNRS-UMR 6203, Avenue Copernic, F-06130 Grasse\\
\email{Olivier.Chesneau@obs-azur.fr}
\and
Department of Astrophysics, American Museum of Natural History, Cent. Park West 79th Street, New York, NY 10024, USA
\and
Max-Planck-Institut für Astronomie, K\"onigstuhl 17, 69117 Heidelberg, Germany
\and
Observatoire de la C\^{o}te d'Azur-CNRS-UMR 6202, Dept. Cassiop\'{e}e,
BP 4229, F-06304 Nice, France
\and
Department of Physics and Astronomy, University of Manchester, Sackville
street, P.O. Box 88, Manchester M60 1QD, UK
\and
Observatoire de Strasbourg, 11 rue de l'Universit\'e, 67000 Strasbourg, France
\and
Department of Physics and Astronomy, Louisiana State University
Baton Rouge, LA 70803, USA}

   \date{Received; accepted }

  \abstract
 {}
   {We present high spatial resolution observations of the dusty core of the Planetary Nebula with Wolf-Rayet central star CPD-56$^\circ$8032, for which indications of a compact disk have been found by HST/SITS observations. }
   {These observations were taken with the mid-infrared interferometer
VLTI/MIDI in imaging mode providing a typical 300\,mas resolution and in interferometric mode using UT2-UT3 47m baseline providing a typical spatial resolution of 20\,mas. We also made use of unpublished HST/ACS images in the F435W and F606W filters.}
   {The visible HST images exhibit a complex multilobal geometry dominated by faint lobes. The farthest structures are located at 7$\arcsec$ from the star. The mid-IR environment of CPD-56$^\circ$8032 is dominated by a compact source, 
barely resolved by a single UT telescope in a 8.7$\mu$m filter ($\Delta \lambda$=1.6$\mu$m, contaminated by PAH
emission). 
The infrared core is almost fully
resolved with the three 40-45m projected baselines ranging from -5$^\circ$ to 51$^\circ$ but smooth oscillating fringes at low level have been detected in spectrally dispersed visibilities. This clear signal is
interpreted in terms of a ring structure which would define the
bright inner rim of the equatorial disk. Geometric models allowed us to derive the main geometrical parameters of the disk. For instance, a reasonably good fit is reached with an achromatic and elliptical truncated Gaussian with a radius of 97$\pm$11AU, an inclination of 28$\pm$7$^\circ$ and a PA for the major axis at 345$^\circ \pm$7$^\circ$.
Furthermore, we performed some radiative transfer modeling aimed at further constraining the geometry and mass content of the disk, by taking into account the MIDI dispersed visibilities, spectra, and the large aperture SED of the source. These models show that the disk is mostly optically thin in the N band and highly flared. As a consequence of the complex flux distribution, an edge-on inclination is not excluded by the data.}
   {}


   \maketitle
%

\section{Introduction}
CPD-56$^\circ$8032 (He~3-1333) is the Wolf-Rayet central star (CSPN) of a young planetary nebula (PN) named PN G332.9-09.9. Wolf-Rayet CSPNe ([WR] stars), which represent about 10\% of all PNe,
are H-deficient stars (mostly [WC] type) that have dense stellar winds and share many characteristics with their higher-mass counterparts from population I (WR stars). CPD-56$^\circ$8032, which is among the coolest CSPNe in this group, is classified as [WC10] (Crowther, De Marco, \& Barlow 1998). There is only a handful of known [WC10] stars and it it must be noted that another member of this small group, He~2-113 (He~3-1044), exhibits striking similarities in terms of spectra, fluxes and nebula extension (i.e. age and distance, D=1.3-1.5\,kpc). An extensive study of the CSPNe of CPD-56$^\circ$8032 and Hen\,2-113 has been conducted by De Marco and collaborators (De Marco, Barlow, Storey 1997 hereafter DBS97; De Marco \& Crowther 1998, hereafter DC98, and De Marco, Storey \& Barlow 1998, hereafter DSB98).
This implies that the two stars and their nebulae have been observed generally with the same instruments
involving similar spectral and spatial resolving power but the lack of spatial resolution prevented De Marco and collaborators from performing an in-depth study of the nebulae whereas their study of central stars  has been exhaustive.

The ISO spectrum of these objects is rich in emission features (Cohen et al. 1999b, 2002).
The Red Rectangle, CPD-56$^\circ$8032 and Hen\,2-113 (Lagadec et al. 2006) are
archetypal polycyclic aromatic hydrocarbon (PAH)-band sources and have amongst the highest ratios of
PAH-band fluxes to total IR flux known. ISO spectroscopy also
found a chemical dichotomy with evidence of crystalline silicates at longer wavelengths (Waters et al. 1998). The initial interpretation of the cold crystalline
silicates was that they probably were located at large distances
from the star, in order to be cool enough. However, the rate of occurrence 
of dual dust chemistry objects is now too high to
be consistent with objects that happened to be caught during
the brief transition from O-rich to C-rich chemistry. Binarity and
the presence of circumstellar disks or tori in which O-rich grains
from a previous evolutionary phase are increasingly
believed to be responsible for the dual dust chemistry (Waters et al. 1988, Cohen et al. 1999b, 2002). 
The binary companion would, in this scenario, form the O-rich disk
during the early AGB and play a role in the heavy mass-loss which
leads to the formation of the [WC] CSPN. 

De Marco \& Soker (2002) proposed that the correlation between Wolf-Rayet
characteristics and dual-dust chemistry points to a common mechanism within a
binary evolution framework (see also Zijlstra et al., 1991). In their first scenario a low-mass main-sequence
star, brown dwarf, or planet spirals into the asymptotic giant branch star,
inducing extra mixing, a chemistry change, and terminating the
asymptotic giant branch evolution. In the second scenario a close binary
companion is responsible for the formation of a disk around either the binary
or the companion. This long-lived disk harbors the O-rich dust. These models
are mostly speculative owing to the rarity of spatial information on the dust
environment at the center of these systems and the scarcity of of binary
companion detection.

Because of the quasi-periodic light variations shown by CPD-56$^\circ$8032, Cohen et al. (2002) suspected the presence of a precessing disk around it. De Marco, Barlow \& Cohen (2002),
presented the first direct evidence for an edge-on disk/torus
around CPD-56$^\circ$8032, revealed by HST/STIS spectroscopy.
Their HST spectra show a spatially resolved continuum split into
two bright peaks separated by 0$\farcs$1 and interpreted to be stellar
light reflected above and below an obscuring dust disk.

The discovery of this disk would probably put this object in the frame of an evolutionary sequence after 
objects like the archetype post-AGB Red Rectangle (Soker 1997, Waters et al. 1998, Men'shchikov et al. 2002) or IRAS 16279-4757 (Matsuura et al. 2004).

Mid-infrared interferometric observations with the MIDI/VLTI instruments are particularly well suited for studying spatially compact dusty environments. We observed
Hen\,2-113 and CPD-56$^\circ$8032 with MIDI in order to constrain the extension of inner regions of their putative disks or rings. Whereas the environment around Hen2-113 appeared sufficiently extended to be fully resolved with the spatial resolution of a single 8m telescope (about 250mas at 10$\mu$m) revealing a 0.5$\arcsec$ wide ring (Lagadec et al. 2006), fringes were obtained from CPD-56$^\circ$8032 demonstrating that the inner parts of the disk are much closer to the star. In addition, the outer environment in the 8.7$\mu$m MIDI acquisition images also appeared well resolved by a single dish telescope, providing further constraints on the dust content of the inner nebula.

These observations, complemented by HST images are described in Sect.2. The images are described in Sect.3 and the MIDI spectra in Sect.4. Sect.5 is devoted to the extraction
of the flux distribution parameters in the N band by means of simple geometrical models. Based on this study, an attempt to derive the physical parameters of the dusty environment by means of more sophisticated radiative-transfer models is presented in Section 6. Finally we discuss the constraints brought by these observations on the geometry and mass of the disk. In particular, we perform a critical comparison of the dusty environment of CPD-56$^\circ$8032 and Hen\,2-113.

\section{Observations}

\subsection{HST}
\begin{figure*}
  \begin{center}
\includegraphics[width=13.cm]{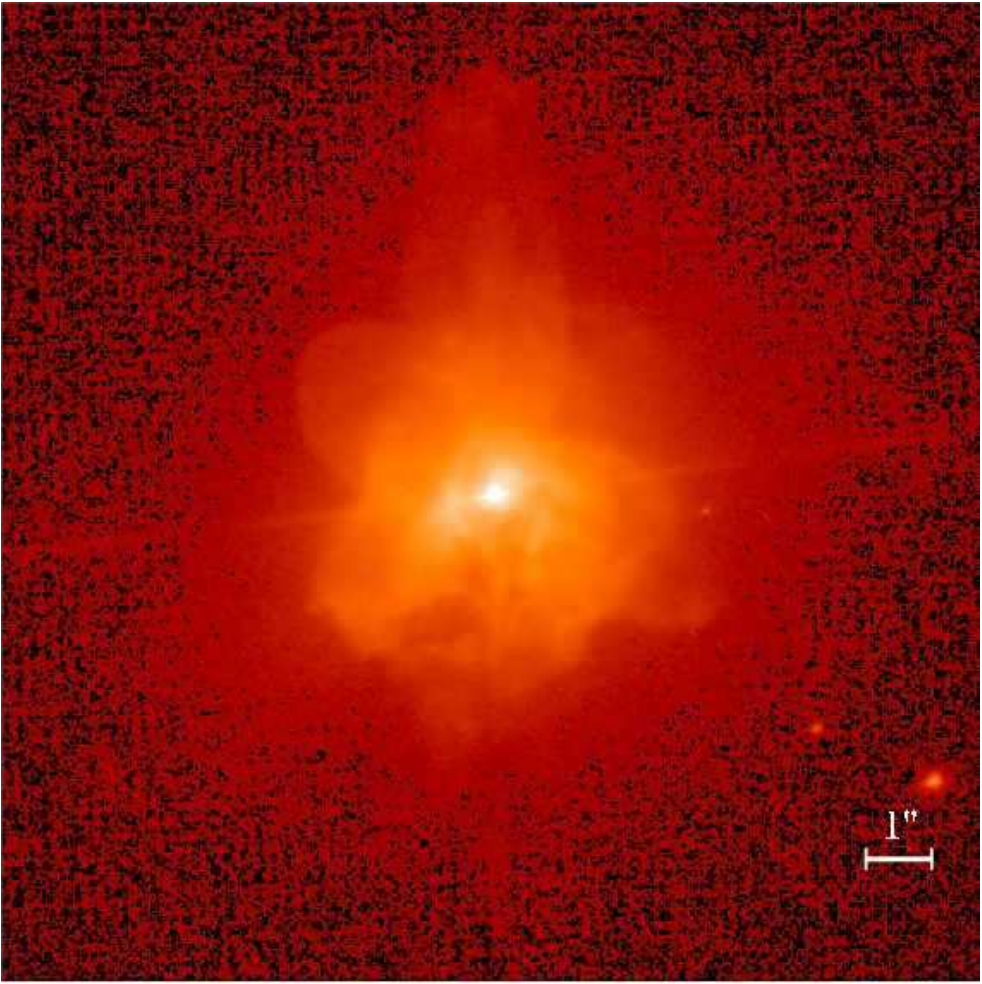}
  \end{center}
 \caption[]{F435W HST/ACS image of the young planetary nebula CPD-56$^\circ$8032 (resolution 0.027$\arcsec$pixel$^{-1}$). A logarithmic stretch has been used. North is up, East is left.
\label{fig:HST1}}
\end{figure*}

Several unpublished exposures of CPD-56$^\circ$8032 taken as part of an HST SNAP imaging program (2002-09-17, GO program 9463, Sahai et al.), were retrieved from the HST archives and processed via the standard calibration pipeline. The data consist of two 225s F435W exposures, and two (34s and 17s) F606W ones. The nebula lies entirely within the field of the ACS/HRC (26\arcsec x 29\arcsec; plate scale = 0.027$\arcsec$ $pixel^{-1}$). 
The F435W combined image is presented in Fig.~\ref{fig:HST1}. It has been rotated and its contrast enhanced in order to emphasize the
faint structures. The F606W images do not differ significantly from the F435 ones, but their dynamic range is very limited.

\subsection{VLTI/MIDI}
The VLTI/MIDI interferometer combined directly the N band light from two VLT Unit Telescopes (UTs), namely the UT2 and the UT3 telescopes. The observations of CPD-56$^\circ$8032 were conducted
during the night of April, 8th 2004 under good atmospheric conditions (clear night, mean seeing of about 0.8$\arcsec$, mean coherence time in visible of 4.6ms). The data were recorded with three different projected baselines, (45.7m,-5$^\circ$), (45.6m,+5$^\circ$),(41.2,+51$^\circ$), hereafter named CPD1, CPD2 and CPD3 respectively.
Custom software developed by the MIDI consortium called MIA was used to reduce
the acquisition images, spectra and fringe data\footnote{http://www.mpia-hd.mpg.de/MIDISOFT/}.
Chopped acquisition images are recorded (f=2Hz, 2000 frames, 4~ms per frame, 98mas per pixel) for the fine acquisition of the target. The default filter is centered at 8.7$\mu$m (1.6$\mu$m wide). 

The journal of the acquisition images is shown in Tab.~\ref{tab:jour} and the journal of interferometric observations in Tab.~\ref{tab:journal2}. The calibrator, HD152786 ($\zeta$\,Ara, N band Uniform Disk diameter of 7.21$\pm$0.21), a K3III star routinely used for VINCI and MIDI observations (Leinert et al. 2004, Richichi \& Percheron 2005, Chesneau et al. 2005) was observed right before or after each science target observations. 

\label{sec:acqim}

\begin{table}[h]
 \vspace{0.1cm}
 \caption{\label{tab:jour} Journal of observations: acquisition images}
 \begin{tabular}{llcrr}
 \hline\hline
 Star & Name & Time &Frames & t$_{exp}$ \\ 
 \hline

CPD-56$^\circ$8032&CPD01&03:46:27 &
1000 & 10s\\
CPD-56$^\circ$8032&CPD02&03:47:12&1000 & 10s\\
CPD-56$^\circ$8032&CPD03&03:48:04&1000 & 10s\\
CPD-56$^\circ$8032&CPD04&03:48:53&1000 & 10s\\
CPD-56$^\circ$8032&CPD05&03:50:09
&1000 & 10s\\
HD152786&PSF01&04:12:39 & 1000 &10s\\
HD152786&PSF02&04:13:41& 1000&10s\\
HD152786&PSF03&04:14:36& 1000&10s\\
HD152786&PSF04&04:16:28& 1000&10s\\
HD152786&PSF05&04:17:50 &1000&10s\\
CPD-56$^\circ$8032&CPD06&05:22:38&1000 & 10s\\
CPD-56$^\circ$8032&CPD07&05:23:21&1000 & 10s\\
CPD-56$^\circ$8032&CPD08&05:24:12& 1000 &10s\\
HD152786&PSF006&05:55:34& 1000&10s\\
HD152786&PSF7&05:56:22 & 1000&10s\\
HD152786&PSF08&05:57:11 &
1000 & 10s\\
HD152786&PSF09&05:57:58&1000 & 10s\\
HD152786&PSF10&05:58:53&1000 & 10s\\
CPD-56$^\circ$8032&CPD09&09:09:13& 1000 &10s\\
CPD-56$^\circ$8032&CPD10&09:10:06& 1000&10s\\
CPD-56$^\circ$8032&CPD11&09:10:55 & 1000&10s\\
HD152786&PSF11&09:28:54& 1000&10s\\
HD152786&PSF12&09:29:45 & 1000&10s\\

 \hline
  \end{tabular}
 \end{table}

We used the interferometric calibrator, HD152786 (K3III) as absolute flux calibrator. For the intrinsic calibrator spectra we used the spectral template from Cohen et al. 1999a of HD163588 (K2III). In interferometry, for each measurement of the source, a measurement of the nearby calibrator is taken and the difference in airmass does not exceed
0.1. Therefore, no airmass correction was applied to the corrected spectra.
Ground-based observations in the N band are difficult
because of the high atmospheric and instrumental background,
and the varying transmission of the atmosphere.
Theoretically, we could have had access to different spectra taken at three different slit orientation of the nebula.
However, the slit was wide (0.6$\arcsec)$ and the variability of the calibrator flux did not allow us to get reliable flux variation which could be attributed to the effect of the slit orientation on the resolved source.
Therefore, the six calibrated spectra taken with two telescopes were averaged and the uncertainty estimated from their statistical variations to about 12\%. 

The extraction of the dispersed visibilities is described in Chesneau et al. (2005b).
The calibrated visibilities are shown in Fig.~\ref{fig:specfringes}. The visibility levels are very low and two error contributions were taken into account: a statistical error of the instrumental measurement computed from the temporal variations of the transfer function measured with calibrators, and a contribution from the systematic uncertainty on the calibrator knowledge and the irreducible level from which the visibility cannot be estimated. While the statistical error is close to zero when the object visibility is very low (as in the present data), the systematic uncertainty is independent of the calibrator measurements and is a measure of the intrinsic noise of the uncalibrated data. This last term, a fundamental limitation to the absolute precision of the visibility measurement, was estimated to a level of 0.005. This value is based on several tests performed on scan selection thresholds of the correlated flux and photometric data of individual measurements of the science target.

\begin{table}[h]
 \caption{\label{tab:journal2}Journal of observations: MIDI/UT2-UT3. The length and position angle of the 
 projected of the baseline is also indicated with their given name.}
\vspace{0.3cm}
\begin{center}
 \begin{tabular}{llcc}
 \hline
  \hline
 Star & Template & Time &Frames \\
 \hline
\multicolumn{4}{c}{09-04-2004, B=45.75m, $\Theta=-5^\circ$, CPD1}\\
  \hline
CPD-56$^\circ$8032& Track. dispersed&03:56:51&8000\\
CPD-56$^\circ$8032& Phot. UT2 disp.&04:03:17&1500\\
CPD-56$^\circ$8032& Phot. UT3 disp.&04:04:37&1500\\
HD152786 & Track. dispersed&04:23:42&8000\\
HD152786 & Track. dispersed&04:27:43&8000\\
HD152786& Phot. UT2 disp.&04:33:52&1500\\
HD152786& Phot. UT3 disp.&04:35:10&1500\\

\hline
\multicolumn{4}{c}{09-04-2004, B=45m, $\Theta=5^\circ$, CPD2}\\
 \hline
CPD-56$^\circ$8032& Track. dispersed&05:30:39&8000\\
CPD-56$^\circ$8032& Track. dispersed&05:34:51&8000\\
CPD-56$^\circ$8032& Phot. UT2 disp.&05:41:26&1500\\
CPD-56$^\circ$8032& Phot. UT3 disp.&05:42:44&1500\\
HD152786 & Track. dispersed&06:04:54&8000\\
HD152786& Phot. UT2 disp.&06:10:57&1500\\
HD152786& Phot. UT3 disp.&06:12:13&1500\\

 \hline

 \hline
\multicolumn{4}{c}{09-04-2004,  B=40.6m, $\Theta=51^\circ$, CPD3}\\
 \hline
CPD-56$^\circ$8032& Track. dispersed&09:16:12&8000\\
CPD-56$^\circ$8032& Phot. UT2 disp.&09:21:33&1500\\
CPD-56$^\circ$8032& Phot. UT3 disp.&09:22:45&1500\\
HD152786 & Track. dispersed&09:35:15&8000\\
HD152786& Phot. UT2 disp.&09:40:57&1500\\
HD152786& Phot. UT3 disp.&09:42:09&1500\\
\hline
  \end{tabular}
\end{center}
 \end{table}

\section{Images}
The nebula around CPD-56$^\circ$8032 is quite faint and up to now only the brightest structures close to the star have been described 
in De Marco et al. (1997).  They can also be seen in the STIS acquisition images (Fig.2 in De Marco et al. 2000).\\
The HST image is complex, in particular several lobes are visible. The general shape is reminiscent of the starfish nebulae as reported by Sahai and collaborators (Sahai \& Trauger 1998, Sahai et al. 2005).\\ 
The most prominent (but with less contrasted) structure is
the north (N) lobe (see Fig.2), which seems to be composed of three distinct ejections. The top one is about 6.85$\arcsec$ long, oriented
with a PA of 9$^\circ$ (from N to E -- the center of angle measurement is the CSPN). These indications are reliable with a
precision of 0.1$\arcsec$ for the angular distance and several degrees (typically 3 to 6$^\circ$), depending on the contrast of the
structure. The second structure is almost as extended as the
first one with 6.2$\arcsec$ but the position angle of 15$^{\circ}$ departs significantly from the first. The third structure seems to be aligned with the first one, but is less extended (3.3$\arcsec$). We note that there is no obvious counterpart of this multiple lobe in the South of CPD-56$^\circ$8032 except a complex filamentary extension at PA of 170$^\circ$.\\
The northeast (NE) lobe, unlike the N lobe, is well defined and very smooth with a dimension of 3.7$\arcsec$ and a PA of 
53$^\circ$. The southwest (SW) lobe could be seen as the counterpart of the NE lobe with the same extension. But, the
tops of the two lobes are not exactly aligned, the SW lobe showing a PA of 240$^\circ$. The fact that the SW lobe is less well defined
than the NE lobe could be due to a perspective effect, the NE lobe being on the front side.\\
Some internal structures are also visible such as two dark lanes, forming a conical structure which points in the  25$^\circ$ E direction. These lanes delimit two areas, the southern one being brighter than the northern one, suggesting a large scale bipolar geometry in which the southern part points toward us (as already suggested in DBS97).
A remarkable bright structure situated east from the CSPN ($\sim0\farcs$5 for the nearest distance to the CSPN), oriented almost N-S, may correspond to 
a bow shock (0.5$\arcsec$x0.4$\arcsec$, B ring in DBS97, see also Fig.2 in De Marco et al. 2002) pointing to the direction at PA$\sim$105$^\circ$. This bow shock is markedly more extended than reported in DBS97, probably because of the increased contrast of our image. Two big holes are also present almost S of the CSPN (0.4$\arcsec$$\times$0.3$\arcsec$ and 0.2$\arcsec$$\times$0.2$\arcsec$).\\
Considering a constant wind velocity of 30\,km\,s$^{-1}$ (DBS97), dynamic ages extend from about $\sim$1400\,yr (N lobe) to 120 years for the bow shock. Note that these structures are also reached by the present wind of the star (225\,km\,s$^{-1}$, DC98) in a relatively short time, i.e. 200\,yr and 15\,yr respectively.  \\

\begin{figure}[h!]
  \begin{center}
\includegraphics[width=7.6cm]{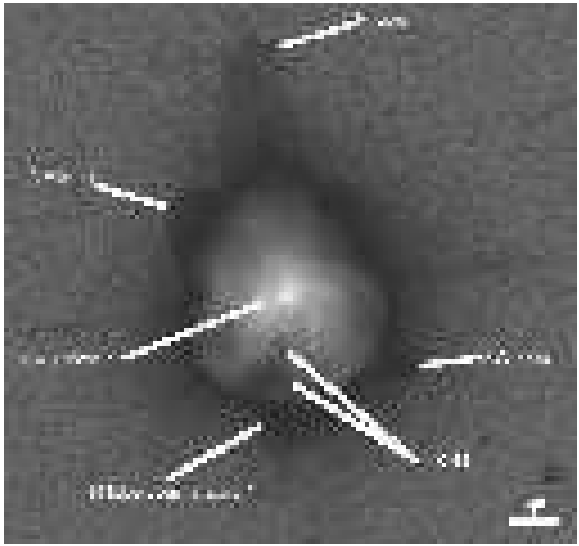}
  \end{center}
 \caption[]{Sketch of the HST F435W nebula. The dynamic scale has been highly exaggerated in order to enhance the remarkable
 structures which are labeled in the figure.
\label{fig:HST2}}
\end{figure}

The MIDI acquisition images reveal an emission more extended than the Point Spread Function (PSF) from a single UT telescope. The extension and geometry of the core was estimated from PSF comparison and Lucy-Richardson deconvolutions with a limited number of iterations (10-20), providing an approximate resolution of about 150~mas (Chesneau et al. 2005a). We used as PSF references most of the acquisition images of HD~152786 (K3III, 82Jy at 12$\mu$m), the interferometric calibrator of the visibility measurements. The observations are recorded
during the acquisition process and the source location within the field
of view can be different for each file. We used the best-centered PSFs which represent 10 acquisitions (over a total of 15). We performed a 2D Gaussian fit for each deconvolved image which provides the image position, extension and the angle of the long axis.
The statistics of the fit yielded a minor axis of $a=289\pm57$mas, a major axis of
$b=429\pm111$mas, an aspect ratio $f=0.68\pm0.07$ and a PA angle of 103.9$^\circ\pm$2.8$^\circ$.
The minor axis is not significantly resolved in comparison to the PSFs extensions taken into account the large error bars.
The PA angle does not correspond to any of the large scale structures seen in the HST image, but is aligned with the bow shock, situated close to the central star at about 1" (see Fig.1).

\begin{table}
\label{tab:resu} \caption{Deconvolved MIDI acquisition images parameters}
\begin{center}
\vspace{0.1cm}
\begin{tabular}{l|c|c}
\hline
Parameter & Mean & RMS\\
\hline
Mean X axis& 289 mas &$\pm$57\\
Mean Y axis&429 mas&$\pm$111\\
Mean ratio& 0.68& $\pm$0.07 \\
Mean PA angle& 103.9$^\circ$ &$\pm$2.8$^\circ$\\
\hline
  \end{tabular}
  \end{center}
 \end{table}

\section{MIDI spectra}
\begin{figure*}
  \begin{center}
\includegraphics[height=7cm, width=8.5cm]{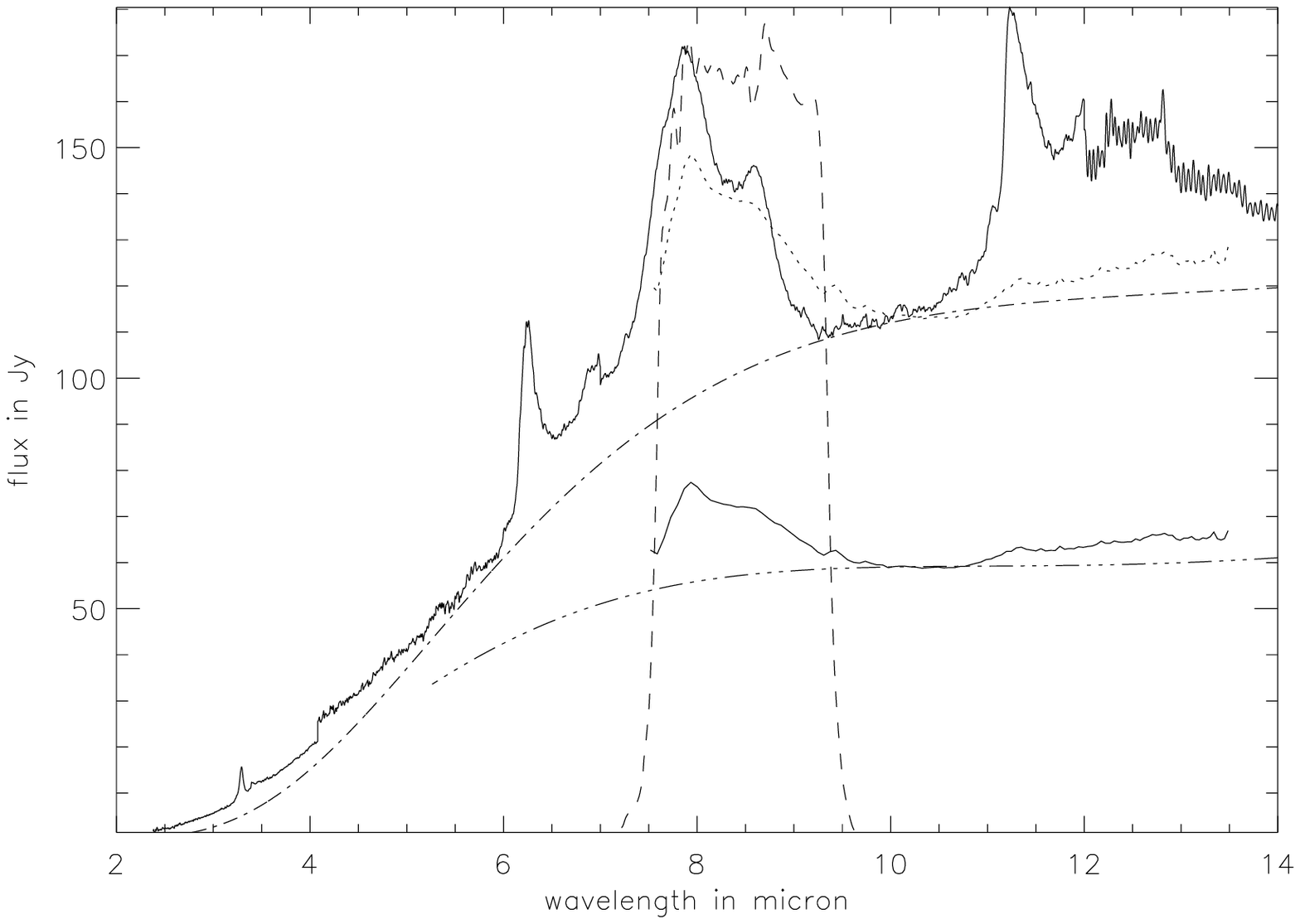}
\includegraphics[height=7cm, width=8.5cm]{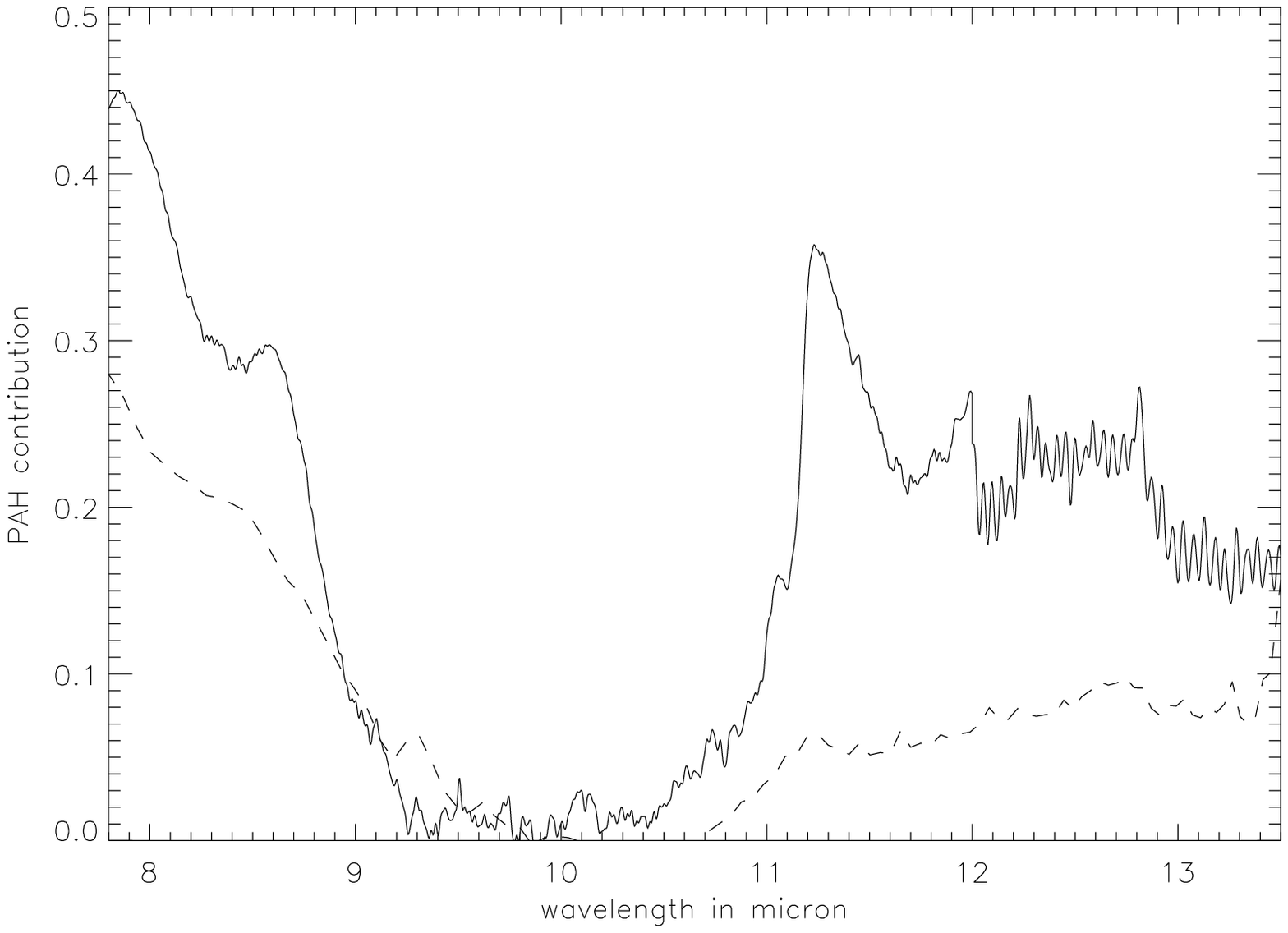}
  \end{center}
 \caption[]{Left: the MIDI spectrum of CPD-56$^\circ$8032 (lower solid) is shown in comparison to the ISO one (upper solid). A MIDI spectrum scaled to the ISO one is also presented (dotted line) to ease the comparison of the PAH features. The transmission of the 8.7$\mu$m filter used for the target acquisition is also shown (dashed). The continuum fit of the ISO spectrum (upper dashed-dotted) is based on the parameters described in Cohen et al. (2002). The fit of the MIDI spectrum (lower dashed-dotted) is based on a slight modification of these parameters (see in text). Right: the result of ratio of the ISO (solid line) and MIDI (dashed line) with their respective approximate continua, showing further the clear discrepancy between the two curves.
\label{fig:spectra}}
\end{figure*}

\begin{figure}
  \begin{center}
\includegraphics[width=5.5cm, angle=-90]{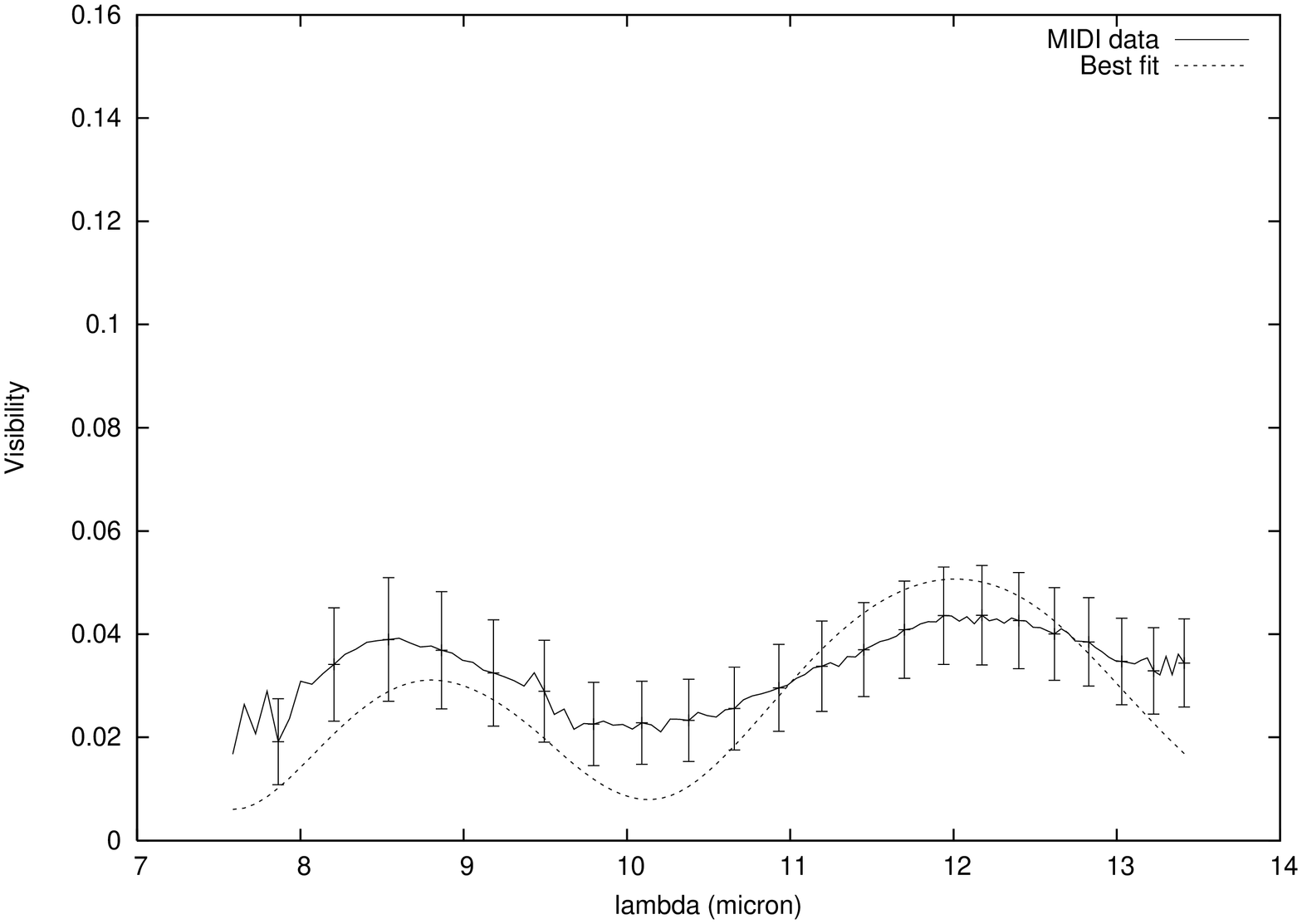}
\includegraphics[width=5.5cm, angle=-90]{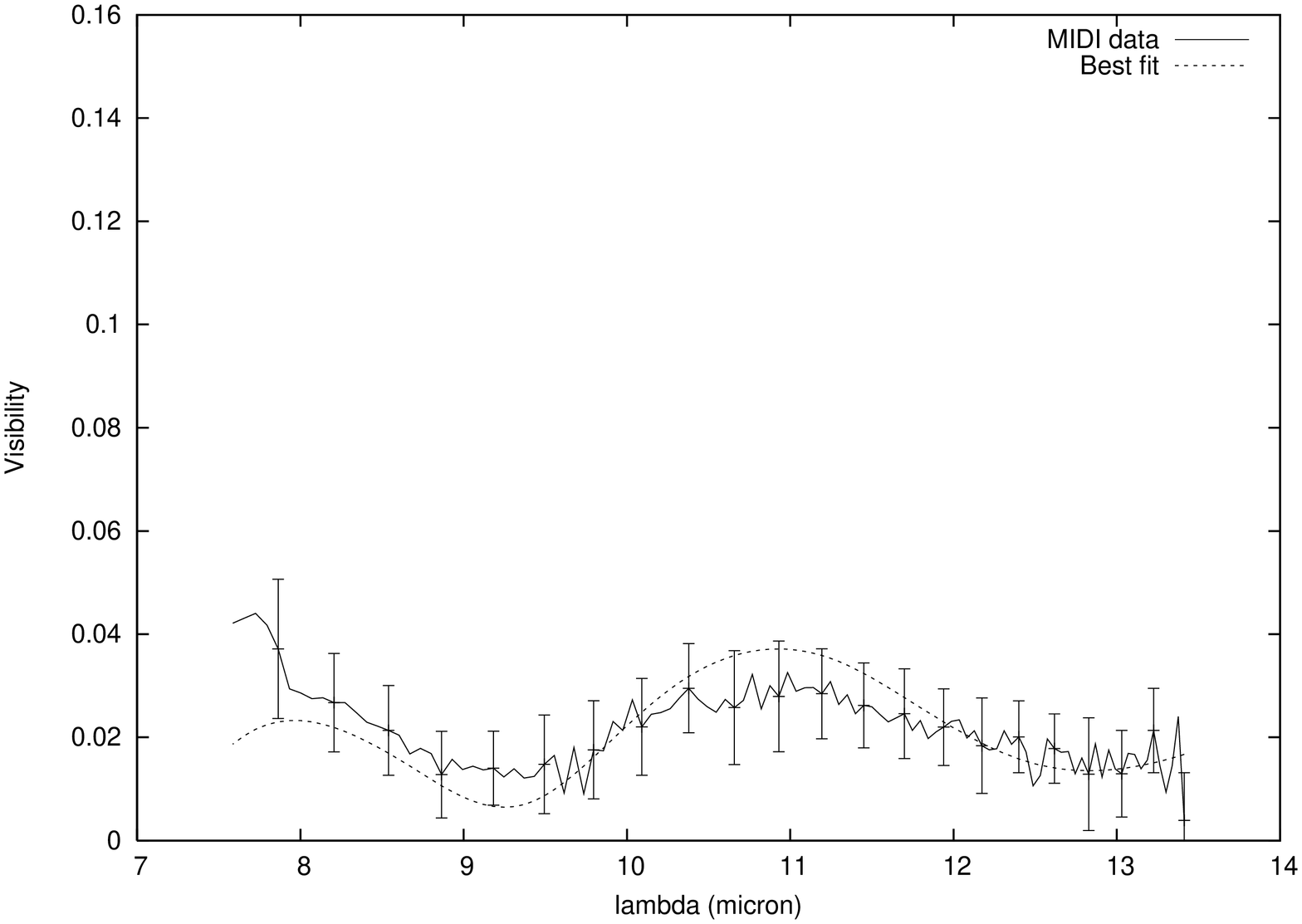}
\includegraphics[width=5.5cm, angle=-90]{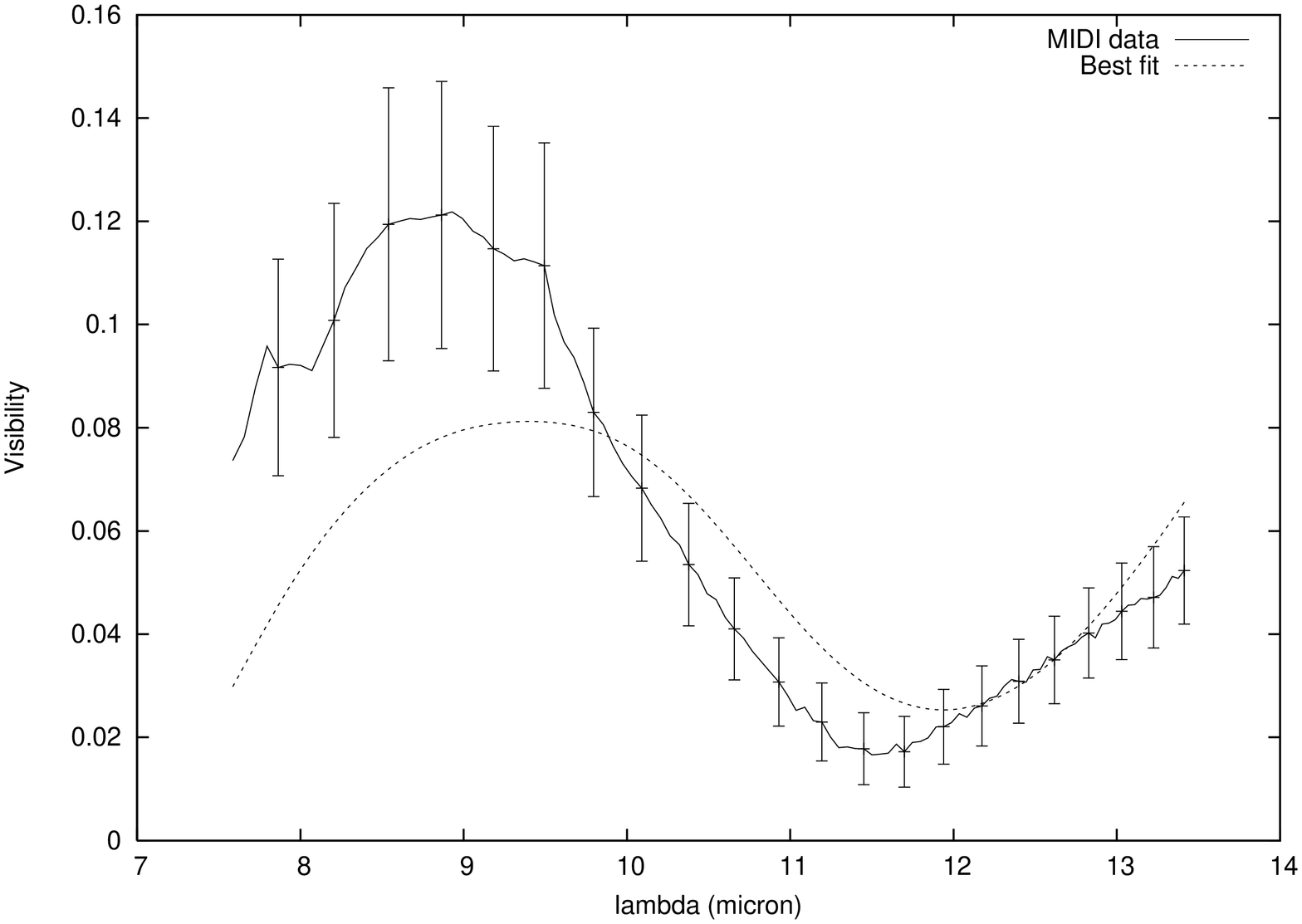}
  \end{center}
 \caption[]{MIDI dispersed visibilities (solid lines, CPD1 to CPD3 from top to bottom) and the 
 best analytical monochromatic models of an inclined disk (Gaussian ring, dotted lines). The $\chi^2$ for the three baselines are 4.3, 1.8 and 14.3 respectively.
\label{fig:specfringes}}
\end{figure}

In Fig.~\ref{fig:spectra} are shown the ISO and MIDI spectra. The ISO spectrum has already been shown and discussed in Cohen et al. 1999b.
The MIDI continuum at 10$\mu$m represents 52$\pm$4\% of the ISO spectrum. The obvious difference between the two spectra is that the 11.2$\mu$m PAH feature is much weaker in the MIDI spectrum when scaled to the ISO one than the 7.9$\mu$m feature which is only slightly weaker. 

The underlying continuum was estimated based on the work of Cohen et al. (2002), who represented the data 
by the superposition of a blackbody having a temperature of 460\,K and that
of a colder gray body ($\lambda^{-0.5}B_{\lambda}(T)$) at 125\,K.

Based on these blackbody parameters scaled to MIDI spectra, we tried to determine a continuum below the MIDI spectrum first by changing the ratio of the global flux, then by increasing slightly the temperature of the hot component to 540\,K in order to account for the change of continuum slope visible in Fig.~\ref{fig:spectra}. Of course, the N band spectrum of MIDI is insufficient to get a good fit for this kind of model and the fit has been adjusted by eye. We evaluated the PAH contribution to about 21$\pm$4\% of the flux in the filter (it would have been 29\% with the ISO spectrum, based on the continuum estimation from Cohen et al. 2002). By comparison, the peak of the PAH fraction in the 11-13$\mu$m region decreases from about 35\% in the ISO spectrum to less than 10$\pm$4\% in the MIDI one. 

We studied also the spatial extension of the MIDI spectra using the technique described in Chesneau et al. (2005). The spectrum is more extended in the PAH bands (7.9 and 11.2$\mu$m) than in the continuum and the extension inferred at 8.7$\mu$m is compatible with the size of the object presented in the previous section. The extensions of CPD-56$^\circ$8032 in the continuum was compared with calibrator ones. The compact core of CPD-56$^\circ$8032 does not seem to be resolved in the continuum. We finally note that we increased the size of the extracting aperture by about 35\% in order to extract the flux from the extended PAHs.

\section{Geometric Models}

In this section are described the analytical fits to the data. 
The goal is to restrain the range of parameters required to account for the data
by means of simple models and physical arguments.

A long baseline interferometer is sensitive to the Fourier transform of the 2D flux distribution from the source.
MIDI operates with a single baseline at a time: this implies that this instrument has a good spatial resolving power in the baseline direction, but only the one provided by a single dish telescope in the perpendicular direction. Hence, a visibility measurement is a measurement of the Fourier transform of the 2D flux integrated perpendicular to the baseline direction, i.e. the Fourier transform of a 1D object (called also strip source). A projected baseline is determined by the alignment of two telescopes as seen from the source, providing an angle and a spatial resolution (proportional to baseline length and inversely proportional to wavelength) that varies with time due to earth rotation. This allows us to recover at least partial information of the object's 2D structure.

The two visibility spectra from CPD1 and CPD2 were obtained with almost identical projected baselines and similar PA angles. The third one, CPD3 was obtained with a slightly lower projected baseline with a clear change of PA angle.
In each case, the object is close to be over-resolved but low level sinusoidal variations are clearly visible. In interferometry, these variations can be attributed to a binary signal or sharp edges in the flux distribution of the object. In the context of a single baseline observation, the binary signal can originate from two single sources or can, for instance, be the result of the 1D integration of a ring-like flux distribution into the baseline direction. The way to disentangle these two models is to observe the evolution of the signal as the projected baseline's PA rotates. 
If the source is a binary, the signal fades as the baseline angle becomes perpendicular to the binary axis. This is not the case for a ring-like structure and we can see in Fig.4 that the signal from the different baselines covering position angles from -5$^\circ$ to 51$^\circ$ is not significantly changed, favoring this geometrical situation. In order to constrain the CPD-56$^\circ$8032 geometry, we tried to match the three baseline dispersed visibilities by means of simple geometrical models. Almost all the models were considered achromatic, i.e. not variable with wavelength, in order to use fully the information provided by the change of resolving power with the wavelength\footnote{The N band is a wide band which shows an increase of 70\% of spatial resolution at 8$\mu$m compared to 13.5$\mu$m}. This is a strong, but necessary assumption considering the small number of measurements available to date. The results from their fits are shown in Tab.~\ref{tab:geom}.

The observed spectral oscillations cannot be the signature of a binary signal of stellar origin. The [WC] central star is too faint to be detected in N band by MIDI (as Hen2-113, the 'twin' of CPD-56$^\circ$8032 in terms of spectral energy distribution (SED) and stellar properties, see Lagadec et al. 2006). Moreover, a putative companion  
bright enough to be detected by MIDI would exhibit a photometric and spectroscopic signature that would have been already detected by existing observations. The disk hypothesis is the most probable in view of the dark lane seen in the HST STIS data (De Marco et al. 2002). Hence, we do not present in the following interpretations based on the assumption that the source can be modeled by a uniform disk or a binary although we note that some tests have been carried out in the checking process. Uniform disks or uniform filled ellipses provide fits of comparable quality to that of the models provided hereafter. These models have been discarded for physical considerations only. The ring-like structure, that could be attributed to the inner rim of this disk, exposed to the radiative flux and wind from the central star, has been chosen as the most physical model. In the case of Hen\,2-113, the structure immediately visible in the MIDI acquisition images was indeed a bright ring embedded in a diffuse nebula extended in the direction perpendicular to the ring main axis (Lagadec et al. 2006). This led us to consider first in priority a circular uniform ring. This geometrical model is defined by two parameters: the inner radius $r_{in}$ and the width of the ring $w$, the flux being constant between the two edges.
The success of the fit is based on the sharp edge of the ring creating the
modulations which mimic the ones detected by MIDI.

We performed first non-linear mean-square minimizations considering each baseline's data individually in order to check the model consistency and detect possible hints of asymmetry. The fits were performed on the 22 spectral channels of the dispersed visibilities considered as independent even though a large part of the error budget on these measurements is correlated, due to photometric fluctuations in the calibration process. In the fitting process, the allowed range of variations for the width of the ring was kept small in order to favor a smooth and relatively slow decrease of the oscillations with increasing spatial frequency rather than a large width leading to a model's visibility closer to the classical uniform disk one. Again, physical arguments favor a steep decrease of the flux as the radius of the disk increases due to stellar flux dilution and the drop of the disk density.

\tiny

\begin{table}[h]
	\centering
		\begin{tabular}{l l l l}
			\multicolumn{4}{c}{{}}\\
			\hline
			\hline
			Parameters& CPD1 & CPD2 & CPD3 \\
			\hline
			\multicolumn{4}{c}{Face-on monochromatic models}\\
			\hline
			Uniform ring : & & &\\
		  Radius $r_{in}$ \tiny{(mas)} & 73$\pm$4&65$\pm$4& 40$\pm$4\\
		  Width $w$ \tiny{(mas)} & 22$\pm$3& 22$\pm$3& 20$\pm$3\\
		  F$_{ring}$/F$_{tot}$	&0.28$\pm$0.05&0.19$\pm$0.05&0.31$\pm$0.05\\
 			$\chi^{2}$/dof &4.2 &3.9 &11.5 \\		 
		  \hline
 			Gaussian ring : & & &\\
 			$r$$_{\frac{1}{2}\rm{FWHM}}$ \tiny{(mas)} & 87$\pm$6 & 91$\pm$6 & 23$\pm$20\\
			$\chi^{2}$/dof & 7.8 & 11.5 & 198 \\		 

		  \hline
		 	\multicolumn{4}{c}{Inclined monochromatic models}\\
			\hline
			Gaussian ring : & & &\\
		  $r$$_{\frac{1}{2}\rm{FWHM}}$ \tiny{(mas)} & 75$\pm$3 & 70$\pm$3 & 73$\pm$3 \\
		  Pos. Angle \tiny{(deg)} & -10$\pm$5 & -10$\pm$5 & -21$\pm$9 \\	
			Incl. $i$ \tiny{(deg)} & 30$\pm$5 & 26$\pm$5 & 29 $\pm$5 \\
			$\chi^{2}$/dof & 4.3 & 1.8 & 14.3 \\		 
		  \hline
		 	\multicolumn{4}{c}{Inclined polychromatic models}\\
			\hline
		  Blackbody model : & & &\\
		  Radius $r$$_{f}$ \tiny{(mas)} & 75 & 70 & 73 \\
		  Pos. Angle$_{f}$ \tiny{(deg)} & -10 & -10 & -21 \\	
			Incl.$_{f}$ $i$ \tiny{(deg)} & 30 & 26 & 29 \\
			T$_{max}$ \tiny{(K)} &1000$\pm$200& 1000$\pm$200 &1000$\pm$200 \\
			$\alpha$ & 2.3$\pm$0.1 & 2.3$\pm$0.1 & 2.3$\pm$0.1 \\
			$\chi^{2}$/dof & 2.5 & 3.4 & 11.8 \\		
		  \hline
		  \multicolumn{4}{l}{$_{f}$: fixed parameter during fitting process}	  
		\end{tabular}
		\caption{\label{tab:geom} Results from geometric model fits.}
\end{table}

\normalsize


The simple two parameter model could not account for the low level of the visibilities so we had to consider a model for which the inner ring flux is diluted into a fully resolved component. The addition of this parameter is not artificial in view of the surrounding complex nebula. This model provides satisfactory fits taking into account the object's complexity already clear in HST images. The different radii found for this model point clearly to an asymmetrical object. Assuming the projection of a circular ring viewed from an inclination $i$ and with the long axis roughly aligned to the -5$^\circ$ baseline CPD1 for which the largest radius was extracted, the three measurements provide an inclination of about 30$^\circ$ (aspect ratio $\sim$0.5). A complementary model consisting of a disk with a bright inner
rim and a rapidly decreasing flux in the outer regions (Gaussian ring) appeared to
provide better results for CPD1 and CPD2 while being unable to fit CPD3 data (the model is not able to account for the rapid increase and the level of visibility close to 8$\mu$m considering the 13.5$\mu$m one). 

In a second step, we performed the fitting procedures with inclined models. Again, these models were compared to each baseline independently and the consistency of the best parameters was checked out. This means that for each model, the main source of information is the visibility change at different wavelengths and not the visibility change for different projected baselines. The Gaussian model was used with an additional parameter, namely the disk inclination $i$. For CPD1 and CPD2 a relatively good fit was reached but the fit of CPD3 is still of worse quality.
However, the consistency of the fitted parameters provided by the baselines is encouraging, the dispersion of the parameters being of the order of the computed internal error.

All of the models considered up to this point are wavelength independent i.e.
the object is supposed to be the same whatever the wavelength of
observation. To allow a kind of wavelength dependence, we also used 
a simple model based on blackbody emission from a disk with a temperature following a power law distribution (T$ \propto T_{in} r^\alpha$).
The inner radii $r_{in}$, angles of orientation and inclination $i$ were fixed and the fit was performed on the parameters $\alpha$ and $T_{in}$. The results show a comforting consistency but the deviations from the data do not
show a dramatic improvement. This suggests that the fits could be more efficiently improved by improving the geometry of flux distributions of monochromatic models than by introducing complex polychromatic ones.

Nevertheless, the quality of the analytical models fits led us to go further into the analysis by using a radiative transfer model.
It must be stressed however that the aim of this new step is limited. We just intend to use the information we got on the object geometry and the radiative transfer model to get some constraints on the physical parameters of the disk, namely the dust temperature at the inner rim, the disk mass and the dust density.


\section{Physical parameters of the disk}
Most of the studies on compact PNe rely on an
interpretation of the observed variable SED by comparing it to a synthetic SED, computed using a
radiative transfer code. In most of the cases, the geometry is poorly constrained and simple geometric configurations
are used. MIDI, however, provides a unique
spectroscopically {\bf and} spatially resolved data set that puts strong constraints on the inner disk geometry and
flux. 


By means of simple radiative transfer calculations with a simple disk model, we try in the following to answer several key questions: what is the approximate mass of the disk? what is the temperature of the disk's inner rim? what is the typical optical depth at 10$\mu$m at the equator of the disk?

We made use of the 3D continuum radiative transfer code MC3D (Wolf 2003; see also Wolf et al. 1999). It is based on the Monte-Carlo method and solves the radiative transfer problem self-consistently. MC3D is designed for the simulation of dust temperatures in arbitrary dust/electron configurations and the resulting observables: spectral energy distributions, wavelength-dependent images and polarization maps.

Below, we briefly outline the strategy used to fit the MIDI observations.

\subsection{A simple disk model}
We use a classical model of a dusty stratified disk (Shakura \& Sunyaev 1973 and Wood et al. 2002).

The dust density follows a 2D law (radial and vertical dimensions):
$$		\rho(r,z) = \rho_{0}(\frac{R_{\star}}{r})^{\alpha} exp[-\frac{1}{2}(\frac{z}{h(r)})^{2}]$$	
where $r$ is the radial distance in the midplane of the disk, $\alpha$ the density parameter in the midplane,
	$R_{\star}$ is the stellar radius and the the disk scale height $h(r)$ is given by $h(r)=h_{0}(\frac{r}{R_{\star}})^{\beta}$ where $h_{0}$ is the scale height at a given radial distance from the star and $\beta$ is the vertical density parameter.\\

We assume the standard interstellar grain size distribution (Mathis et al. 1977): $\frac{dn(a)}{da}\  \alpha\  a^{-3.5}$ where $a$ is the dust grain radius. Grain radii extend from $0.005$ to $0.25 \mu$m (Draine \& Lee 1984). We consider the dust grains to be homogeneous spheres.
	
The dust is in the form of graphite grains. Graphite has been used to model the dust in PN for instance in 	
Stasi\'nska et al. 	2004, Stasin\'ska, \& Szczerba, 2001), with a similar $\alpha$ parameter.

The stellar parameters of CPD-56$^\circ$8032 are relatively well constrained (De Marco et al. $1997$) and these are summarized in Table~\ref{tab:modparam}. We also report in this table the constraints on the disk geometry inferred in the previous section.

\begin{table}
	\centering
		\begin{tabular}{cc}
		\hline
		$T_{eff}$ (K) & $30000$\\
		Luminosity $(L_{o})$ & $5100$ \\
		Distance (kpc)& $1.35$\\
		\hline
		Inner radius (AU) & $97\pm11$ \\
		Inclination ($^{\circ}$) & $28\pm7$ \\
		PA angle ($^{\circ}$)& -15$\pm$5\\
		\hline
		 
		\end{tabular}
	\caption{Model parameters based on DBS97, and from the geometric fitting using a truncated Gaussian.}
	\label{tab:modparam}
\end{table}

Considering the complex geometry, witnessed by the HST images, and the complex chemistry exhibited by the ISO spectra, we did not attempt to develop a fully consistent model for CPD-56$^\circ$8032 and its environment.
We concentrated our efforts on several particular points: 
\begin{itemize}
\item The ISO continuum emission was considered as an upper limit to the model emission. The outer radius of the model was chosen arbitrarily to have a size of 3000\,AU (2$\farcs$2 at D=1.35\,pc). 
\item We did not try to model the MIDI acquisition image at 8.7$\mu$m since we believe that the asymmetry is probably the consequence of the PAH distribution alone (see discussion in Sect.7.2). However, we performed consistency checks in order to ascertain that the continuum images from the model are not more extended than the MIDI one. 
\item The MIDI spectrum was taken into account as an important constraint for the N band flux originating from the inner regions of the disk. Again, due to the PAH contamination, we compared only the global spectrum slope and the level of continuum.
\item We tried to match the MIDI visibility curves as closely as possible. 
\end{itemize}

The parameter space chosen was restricted to the ranges provided by the geometric fits (see Table~\ref{tab:modparam}).

\begin{table}
	\centering
		\begin{tabular}{ccc}
			\hline
 			  Composition & Graphite & Graphite/Silicate\\
 			\hline
 			  R$_{out}$ (AU) &  $3000$  &   $3000$ \\
 			  R$_{in}$ (AU) & $105$     & $105$ \\
 			  R$_{dis}$ (AU) & - &    $1000$\\
 			  i &  $21^{\circ}$ &   $24^{\circ}$\\
 			  PA & $-15^{\circ}$ &  $-13^{\circ}$\\
 			 \hline
 			  $\alpha$ & $2.0$ &  $2.1$\\
 			  $\beta$ & $1.5$ & $1.3$\\
 			  h$_{100AU}$ (AU) & $12$ &  $11$\\
 			  M$_{dust}$ (M$_{\odot}$) & $2.10^{-4}$    & $7.5.10^{-4}$ \\
 			  m & - &  $2.75$\\
 			  T(R) & 518 K & 606 K\\
 			  \hline
		\end{tabular}
		\caption{Parameters of the best models.}
		\label{tab:recapparam}
\end{table}







The outputs from the code are, a large aperture SED, 15 images with pixel size of 13mas in the wavelength range 8-13.5~$\mu$m. By applying a Gaussian aperture (from 210mas at 8$\mu$m to 350mas at 13.5$\mu$m) to the images, we are able to extract the equivalent of the MIDI spectrum and the spectrally dispersed visibilities (the limited spectral resolution is taken into account).

To transform the images into a visibility signal, the images are first rotated by the PA of the object in the sky and then collapsed as 1D flux distributions in the direction of the three baselines. The visibility for each wavelength is the value of the Fourier transform of these 1D vectors, normalized to the zero frequency value at the considered projected baseline. 



\subsection{Results}

\begin{figure}
  \begin{center}
\includegraphics[height=8cm, width=5cm, angle=-90]{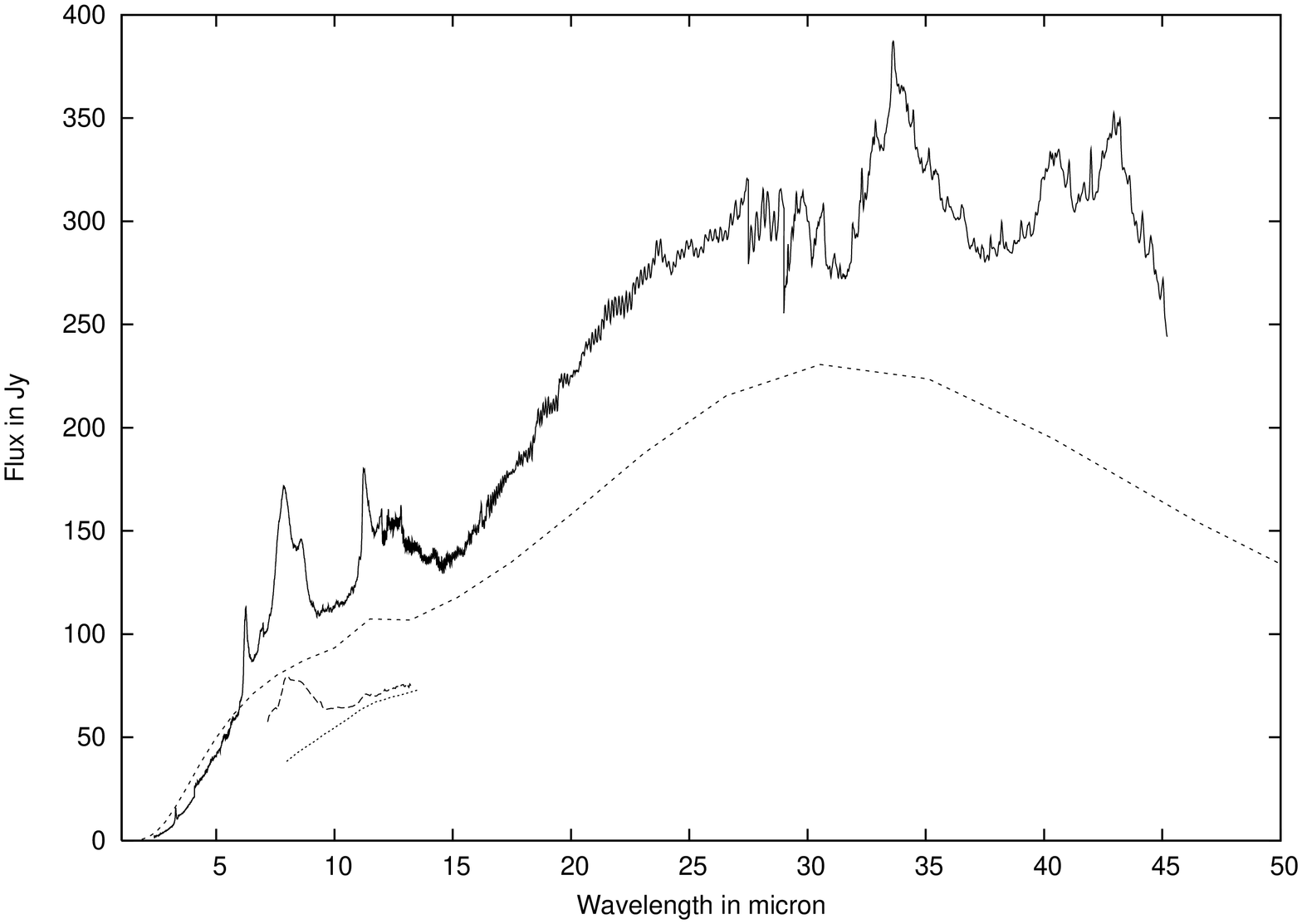}
\includegraphics[width=5.5cm, angle=-90]{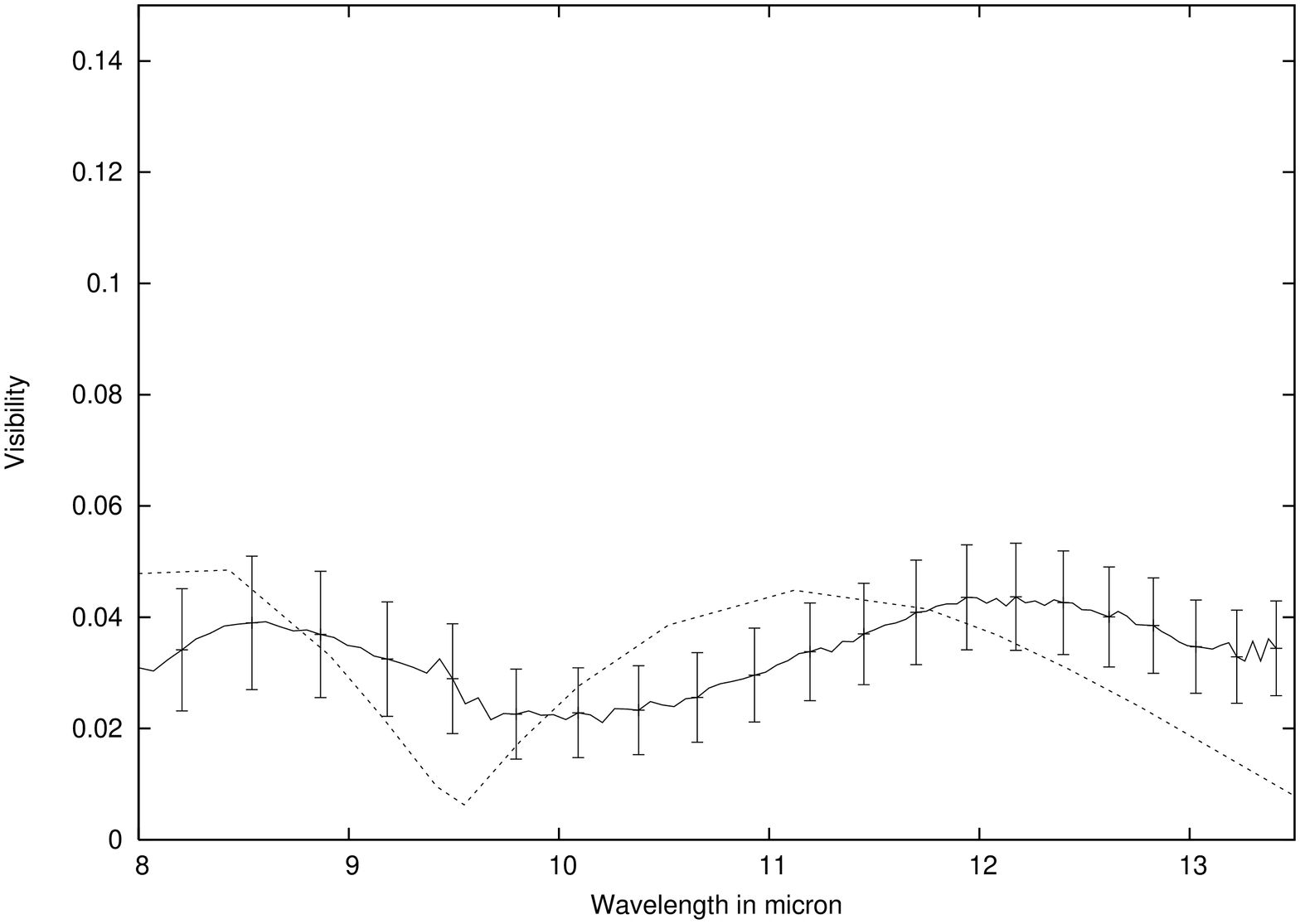}
\includegraphics[width=5.5cm, angle=-90]{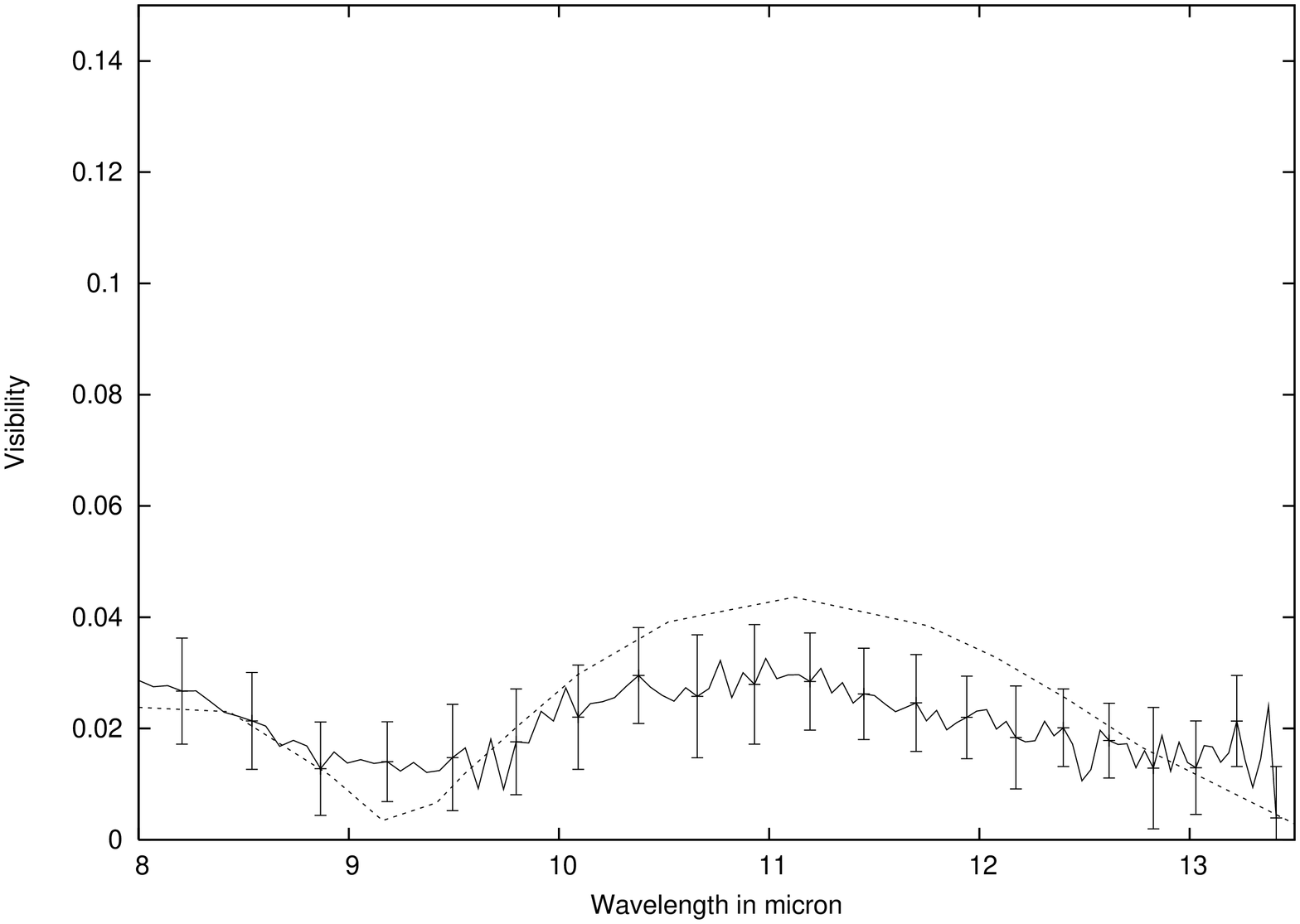}
\includegraphics[width=5.5cm, angle=-90]{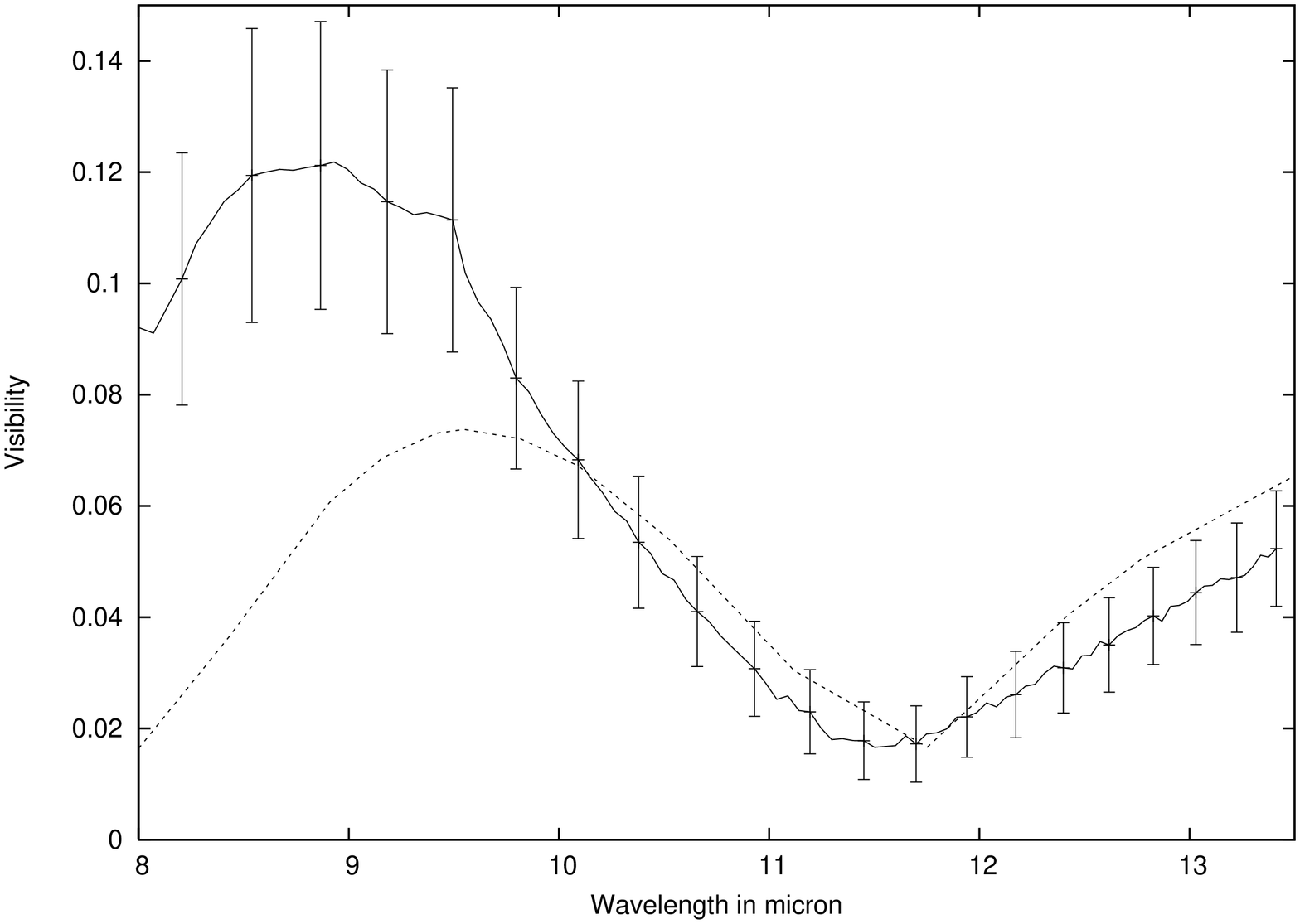}

 \end{center}
 \caption[]{One of the best models with single chemistry. Top: the ISO spectrum (solid line) is compared with the model SED (dotted line). The MIDI flux (dashed line) is very close to the model one (within 10\%, small dotted line). The $\chi^2$ for the three baselines are 7.5, 3.9 and 10.4 respectively.
\label{fig:dust1model}}
\end{figure}

In Fig.~\ref{fig:dust1model} and in Table~\ref{tab:recapparam} we show good model fits, and parameters, respectively. The first models were able to provide a reasonable fit of the SED with the vertical density parameter $\beta$ with its classical value of 1.25. However, in order to account for the MIDI flux, it was necessary to increase the amount of hot material close to the star. This was first carried out by setting $\alpha$ to 2.3-2.5, but the dispersed visibilities were much higher than the observed ones suggesting that the 10$\mu$m flux was too dominated by the inner regions. Moreover, the slope of the continuum extracted from the images using an aperture equivalent to the one of a 8m telescope did not correspond to the MIDI one. Decreasing the disk mass or increasing the scale height $h_{0}$ of the disk was helpful but not sufficient to solve the discrepancy. Finally, increasing the parameter $\beta$ of the disk (the 'flaring') appeared to be the most efficient way to reconcile the MIDI spectrum and the visibilities. A consistent model was found by working mainly with both the scale height and the parameter $\beta$ of the disk. The final model is a flared disk which is optically thin from almost any direction (the equatorial plane optical depth at 10$\mu$m is close to 1).
  
We also tested models with additional parameters, namely the radius at which a density discontinuity R$_{dis}$ occurs and a multiplying factor $m$ for the density at the discontinuity radius. The exponent of the density law, $\alpha$, remains unchanged. In these models silicates replace graphite in the outer part of the disk.
Several values of the R$_{dis}$  were tested, and a value of 1000\,AU ($\pm$100\,AU) provided a fit as satisfactory as the single chemistry model. This could be explained by a modification of the mass loss regime about 200 years ago, considering a constant wind velocity of 225\,km\,s$^{-1}$. These models do not affect much the MIDI visibilities, which are more sensitive to the inner regions of the disk. The discontinuity would be located at about 0.6-0.8\,arcsec from the star, i.e. in a region observable with a single 8m telescope in the N band with deep exposures. We note that this value of R$_{dis}$ is in agreement with the Cohen et al. (1999b) findings. 
The disk is optically thin, changing the flaring or the scale height parameters modifies the appearance of the object in a complex manner and the inclination is no longer a dominant parameter of the fitting process. We tested many models at inclination close to 90$^\circ$, first because the HST observations from De Marco et al. (2002) suggest it, second because an edge-on optically thin ring is seen as two bright spots that can generate visibility oscillations as observed in our data. The period of these oscillations is at a minimum when the baseline is aligned with the disk PA. As the baseline angle gets close to perpendicular of the disk, the projected spot separation decreases leading to an increase of the period of the visibility oscillations. The increase of about 40\% observed between the CPD1 and CP3 visibilities is compatible with this scenario. Yet, the scale height of the disk is a non-negligible fraction of the disk inner radius and the baseline CPD3 that mostly constrain the system inclination also probes the vertical extent of the disk. We never reached the same quality of visibility fits as for the low inclination models although the parameters space has not been fully explored.

To conclude, the radiative transfer models, whose parameters are closely related to the ones extracted from simple geometrical models appear to give good results provided that the disk is optically thin in N band and highly flared. Viable models were found at inclination lower than the 28 degrees from the geometrical models, but, owing to the large vertical extent of the disk, we stress that a close to edge-on inclination are not excluded. Definitely, CPD-56$^\circ$8032 is a complex system and many more baseline observations are needed to constrain any detailed model.


\section{Discussion}
\subsection{Geometry and the mass of the disk}
The dispersed visibilities provided by MIDI have allowed us to put stringent constraints on the spatial distribution of the
N band flux in the close vicinity of the central star. A ring structure has been discovered with an inner radius of about 97\,AU (using D=1.35\,kpc) and a PA of the major axis of about -15$^\circ$. The PA is close to perpendicular to the slit direction of the STIS/HST observations of De Marco et al. (2002) at PA=63$^\circ$, which confirm the suggestion from the authors that their slit was fortunately oriented to detect the disk.
The inner rim temperature of the disk is in the range of 500-600\,K and the inner regions provide the flux interpreted by Cohen et al. (1999b, 2002) as the warm component. 

The 0$\farcs$10 angular separation between the two main scattered light
components in the STIS spectral image were assumed by De Marco et al. to correspond
to the top and bottom of an edge-on disk or torus with a
projected disk semi-thickness of 67 AU (for a distance of
1.35 kpc; De Marco et al. 1997). Our models provide an inclination that differs significantly from their assumptions. Nevertheless, the scale height of the best model disk at the location of the inner rim ($\sim$100\,AU) is 10\,AU and taking into account the value of the disk opening angle ($\beta$=$\sim$1.4), and the inclination, we find a projected thickness of the disk compatible with the estimate of De Marco et al. In the case of a close to edge-on configuration, a significant part of the observed visible flux from the central star should be scattered in the flared borders of the disk whereas in the low inclination case, most of the flux is directly sent toward the earth. We note that the measured reddening for CPD-56$^\circ$8032 in De Marco et al. (1997) is relatively low (E(B-V)=0.68), even lower for instance than the one inferred for Hen\,2-113 (E(B-V)=1.0). Moreover, the visual magnitude of CPD-56$^\circ$8032 and Hen2-113 are 10.9 and 12.3, respectively for similar estimated distance, suggesting that the circumstellar visual extinction of the stellar flux is small compared to Hen2-113. A perfectly edge-on opaque disk in which the scattered star light emerges from the upper and lower rims of the disk would probably imply a much lower visual magnitude. In the context of the low inclination hypothesis, the presence of many PAHs grains in the line of sight could account for a part of the UV absorption seen by the HST without affecting the visual flux of CPD-56$^\circ$8032 very much. 

The option of an edge-on disk was favored by Cohen et al. 2002, mainly because three visual light declines were observed having an apparent periodicity of $\sim$5 yr (Cohen et al. 2002, Pollacco et al. 1992). Cohen et al. suggested that these light declines could be due to dense clumps orbiting within the
disk, in which case the apparent periodicity might be accidental. A fourth decline happened at the end of 2004, earlier than would be expected by a 5\,yr periodicity since the previous event occurred in May 2001 (Albert Jones, private communication).

CPD-56$^\circ$8032 is the archetype of sources exhibiting double chemistry. However, we stress that MIDI operates in the N band and probes only the hot carbon rich content of the disk. There is definitely no trace of any silicate feature in the correlated flux detected by MIDI. The total mass of carbon dust implied in our models is limited to 2-4\,x\,10$^{-4}$M$_\odot$ (without considering any PAH grains). A rough estimate of the oxygen crystalline content from Cohen et al. (1999b, 2002), based on the ISO spectrum longward 25\,$\mu$m is about 3.8\,x\,10$^{-4}$M$_\odot$, which could represent about 40\% of the total silicate content of $\sim$ 1\,x\,10$^{-3}$M$_\odot$. 
The flared and optically thin carbon disk of CPD-56$^\circ$8032 suggest that an efficient dissipating process is occurring. This interpretation is compatible with the suggestion of Cohen et al. (1999, 2002) that the recent wind and increase of UV flux from CPD-56$^\circ$8032 affects deeply a preexisting cold and probably equatorially enhanced Oort cloud that could be the relic of a dense long-lived oxygen-rich disk generated during the AGB phase.

\subsection{PAHs}
Based on the data in hand, it is difficult to get a real idea of the spatial distribution of the PAHs around CPD-56$^\circ$8032.
Nevertheless, we have noted several intriguing results:
\begin{itemize}
\item The 8.7$\mu$m acquisition image of MIDI is probably affected by PAHs, but it is difficult to prove it in the absence of any continuum reference image. The direction of the elongation at 103$^\circ$ does not correspond to the direction perpendicular to the disk main axis as determined in this study, but rather coincides with the bow shock at about 0$\farcs$5 from the star visible in the HST images (this study and De Marco 1997, 2001). In the case of CPD-56$^\circ$8032, the complex mass-loss history may lead to preferential PAH emitting regions at the interface of young outflows and older structures. Due to the absence of hydrogen in the wind, PAHs cannot be present there and are created only when hydrogen-deficient soots penetrate into the H-rich nebula. This means that the PAHs are probably found preferentially at the disk interfaces in the vicinity of the star and in the contact discontinuities at larger distances. For the Red Rectangle, a geometrically well-defined object, the complexity of the chemical environment is striking (Markwick-Kemper et al. 2005).
\item A strong 7.9$\mu$m band and a weaker 11-14$\mu$m plateau are observable in the MIDI spectra whereas the balance of the flux between these two bands in the ISO spectrum is markedly different (see Fig. 6). The PAH bands from 7 to 9$\mu$m (C-C modes) behave generally in a decoupled way compared to the 3$\mu$m and 11-14$\mu$m plateau (out-of-plane C-H modes). This plateau is attributed to a blend of many bands due to
deformation modes of large and {\it mostly neutral} PAH molecules (Hony et al. 2001, Matsuura et al. 2004). In particular Joblin et al. (1996) interpreted the variations of the 8.6 and 11.3$\mu$m bands in terms of ionisation state effects on the CH bending mode intensities, the PAHs being positively charged in regions of high UV excitation. This favors the interpretation that the PAHs seen in the MIDI spectra are mostly cations which directly face the wind of the Wolf-Rayet star.
\item The observed PAH features seen in the MIDI spectra do not seem to imprint their mark in the visibility curves. This can be interpreted in two ways. The first hypothesis is that the PAHs follow exactly the same geometry as the carbon-rich disk and that carbon and PAHs are perfectly mixed. In this frame, it is difficult to understand why the 8.7$\mu$m image should be elongated in a particular direction, not related to the disk geometry, in contradiction to the evidence presented above. Moreover, PAH bands are only seen in relatively unshielded
(low-extinction) gas and it is difficult to imagine why their emission mechanisms would lead to the same flux distribution as the continuum from amorphous carbon. The second and more probable explanation is that the PAHs mostly emit out of the 8m telescope beam and do not contribute to the visibility pattern extracted from a smaller aperture than the MIDI spectra. We have checked carefully the two apertures and the difference is significant (of the order of 30\% in FWHM) and sufficient to account for the spectral variations. 
\end{itemize}

\subsection{CPD-56$^\circ$8032 and Hen2-113: two twins with different nebulae}
\label{sec:comCPDHEN}

Hen\,2-113 and CPD-56$^\circ$8032 offer a unique opportunity to compare two different nebulae ejected from central stars presenting striking similarities, especially in terms of the distance estimation. This implies that the two stars and their nebulae have been observed generally with the same instruments involving similar spectral and spatial resolving power.
On a large scale, their environments are quite different. The nebula around Hen\,2-113 is slightly more compact and probably more simple than the one seen around CPD-56$^\circ$8032. Hen\,2-113 shows a 1$\arcsec$ ring-like structure which is well resolved with a 8m telescope from the L' to N bands, whereas CPD-56$^\circ$8032 harbors a compact disk in the vicinity of the central disk. Could these differences only be explained assuming that their dynamical ages are markedly different?

If we assume a similar dynamical age, can the general appearance of these nebula be a consequence of binary interaction? In the case of CPD-56$^\circ$8032 we could answer that the probability is strong but in the case of Hen\,2-113, the multi-lobal geometry is the only convincing evidence. There is a correlated issue: Lagadec et al. (2006) have shown that the bright core of Hen\,2-113 in the L' and M' bands consists in an optically thin shell of newly formed dust that surrounds the star. A question to be answered is how this dust is formed in the hostile environment of hot winds. 
Dust formation (in particular episodic dust formation) might be the result
of binary intereaction as is the case of the massive stars WR104 and WR98a
(Tutthill et al. 1999; Monnier et al. 1999, 2000). The presence of
circumstellar or circumbinary disks is another signature of the possible
presence of an unseen companion orbiting CPD and He2-113.

In the case of [WR] CSPNe, De Marco \& Soker (2002) put forward the
speculative hypothesis that they result from a merger which dictated the
change between oxygen-rich and carbon-rich chemistries as well as a
departure from the AGB. In this scenario, the stars are currently single
(we note that no [WC] CSPN is known to be in a binary system) and the
disks would be a signature of their binary past. In this scenario the newly formed dust seen around Hen2-113 
(Lagadec et al. 2006) would not be triggered by a companion.

\section{Conclusion}
We have presented new high spatial resolution observations of the dusty environment of the PN CPD-56$^\circ$8032, from HST/ACS and MIDI/VLTI instruments.
These observations confirm that a compact disk is present at the center of the PN with a position angle of its projected major axis close to the one inferred by De Marco et al. (2002). The signature of the inner rim of the disk is quite clear, evidenced by the oscillations detected in the MIDI dispersed visibilities and the radius of this rim has been determined to be of the order 75\,mas that, at a distance of 1.35kpc translates to about 100\,AU. The recorded MIDI continuum flux is about half of the ISO flux and we have shown that in the 8.7$\mu$m filter, the compact core is resolved by a single dish 8.2\,m telescope.

The parameters extracted in this paper have used a very limited sample of interferometric measurements taken in the context of testing the feasibility of observing this kind of object with MIDI, with the help of a complex HST image taken in visible light and a well-constrained large-aperture SED thanks to the ISO spectrum. In particular, the inclination of the disk is not uniquely constrained by these observations.
The next natural step is, as for Hen2-113, to observe CPD-56$^\circ$8032 in the K, L and M bands with the VLT/NACO adaptive optics instrument, which is able to provide images at the scale of 100\,mas. The bright disk (F$_{12\mu m}$=143\,Jy) can also be studied in depth with MIDI thanks to the new 1.8m VLT Auxiliary Telescopes, providing a great wealth of short (8-32m) baselines although we note that the fringes from this extended object will be recorded from a much larger beam. However, the present study has shown that most of the flux originated from regions unresolved by a 8m telescope, and this problem should easily be solved by the large number of baselines expected for these observations.


\begin{acknowledgements}
We thank N. Nardetto and A. Meillant for help and advice.

\end{acknowledgements}

\end{document}